# A theorem on the photographic process in Special Relativity
## The train paradox revisited

**M Azreg-Aïnou**

Başkent University, Engineering Faculty, Bağlıca Campus, Eskişehir Yolu 20 km, Ankara, Turkey[1]

**Abstract**
The purpose of this letter is to show, on the one hand, how the so-called train paradox could be resolved directly without appealing to non-linear Lorentz transformations. The resolution is established in the most general case of curvilinear motion with a variable speed train. On the other hand, we formulate a theorem on the photographic process of two moving objects with relativistic effects.

## 1. Introduction

In the Special Theory of Relativity, measurements and observations are dependent on the referential frame and remain relative to it; nonetheless, two facts remain absolute and independent of the referential frame: 1) the constancy of the speed of light in vacuum and its independence of the speed of the source, which is the second postulate of Einstein, and 2) the coincidence of two events[2]. Any other absolute fact encountered in the Theory is a mere consequence of the second postulate.

An important feature of the postulates of Special Relativity is that one can use them directly to solve a variety of problems by theoretical treatments not appealing explicitly to Lorentz transformations. Einstein himself and some other authors [2,3] applied this method, and we shall do this[3]. For instance, problems involving light propagation are better understood when first solved by applying the second postulate of the Special Theory of Relativity. Resolving them afterwards by applying Lorentz transformations would be a convincing approach regarding the teaching of the Special Theory of Relativity. This is because the latter has brought to us new ideas and is based on a new logic that is different from our old classical one. To be admitted intuitively, this new logic—as any new one, say, for instance, Quantum Mechanics—should be practiced and experienced by different approaches and from different viewpoints.

That said, the way the paper is organized in not as its title indicates; the approach is totally reversed. Motivated by the issues which have arisen in [4,5] and the ones recently resolved [1], we will explain in section 2 how the train paradox arose by misapplication of Special Relativity to the case of two objects in relative motion. We will resolve it simply by applying the two absolute facts of Special Relativity mentioned above in the first paragraph of this introduction. We will end the section by commenting on [1]. This resolution will help us in writing the definitions and formulating the assertions of a theorem on the photographic process of moving objects. The theorem, as well as the proofs and comments, is the matter of Section 3. Our aim is to clarify the issues related with photographing moving objects in order to achieve a holistic understanding of the problem and to avoid illusions that could lead to paradoxes, and to supplement our pedagogy. Our conclusions are summarized in Section 4.

---

[1] E-mail: azreg@hermes.unice.fr
[2] This fact is valid in any relativity theory and does not belong to Special Relativity, in the sense that it does not originate in the Theory.
[3] Our remark regarding the paper [2] is the following. We read on page 69 of [2]: "*Of particular interest is the role of the Special Theory of Relativity: the question of when it is relevant to cases of superluminal motion…*". In Section 3 of [2], the authors derived their formulas by supposing that the speed of light is constant and independent of the speed of the source, which is a postulate, and a fundamental postulate, of Special Relativity. The Special Theory of Relativity should not be envisaged as the set of Lorentz transformations for these are derived from, and hence are the consequences of, the second postulate and other hypotheses.

## 2. The train paradox

In Special Relativity, the notions of measuring the size of an object and photographing it are completely different. By definition, measuring means considering all or a set of points that built the object at the same date (instance). While photographing or watching an object, even if it is at rest, means considering all or a set of points that built the object or just its envelope at different dates. Hence, the result of measuring a uniformly moving sphere is an ellipsoid due to Lorentz contraction parallel to the velocity vector, while the result of photographing a uniformly moving sphere is, under some conditions, a circular disk[4] that is the projection of the sphere onto the film [6].

The consideration of the photography of two objects, in relative motion with respect to each other, calls for a careful treatment. The two objects must be treated together and not separately. The train paradox arose because of such a rough treatment [4-5], as is the case with most of the paradoxes of the theory. It is roughly admitted that an extended moving object appears to be rotated on the film due to its motion and to the effect of signal delay. Hence, the paradox: the moving train undergoing the rotation appears off the rails, at rest, in the referential frame of observation. Let us call longitudinal sections those lines crossing the object parallel to its translation velocity vector. As we shall see below, the true assertion is such that those non-longitudinal sections appear having rotated with respect to the longitudinal ones on the printed film, with an angle of rotation depending on the speed of the object. The stronger assertion is that the longitudinal sections do not undergo any rotation that could be due to the speed of translation of the extended object.

Consider a train moving with a changing velocity vector with respect to an inertial frame $S$. The rail, at rest in $S$, is a smooth curve $y = g(x)$ in the plane $z = C < 0$, as shown in Figure. The point of observation is at rest at the origin $O$ of $S$, and the plane (P): $y = D > 0$ represents the film, with the optic axis of the observer in the positive y-direction. Let $(U,V)$ be the coordinates of a point in (P), with the U- and V-axes parallel to the x- and y-axes, respectively, and the origin of (P) on the y-axis.

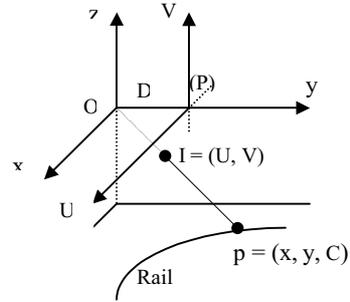

**Figure.** When the point p traces the curve $y = g(x)$, $z = C$, the point-image I traces the curve $V\,g(CU/V) = DC$ in the plane (P).

We consider the light rays from the train and the rail that reach $O$ at $t = 0$. We will first find the shape of the image of the rail on (P). Consider a particle $p$ of coordinates $(x,y,z)$. The light ray from $p$, at $t = -\sqrt{x^2 + y^2 + z^2}/c$, will impress the film at a point $(U,V)$ given by [1] $(U,V) = (Dx/y, DC/y)$. Solving for $(x,y)$ yields

$$(x, y) = (CU/V, DC/V). \qquad (1)$$

For a particle on the rail ($p_R$), this writes $(x_R, y_R) = (CU_R/V_R, DC/V_R)$. Since $(x_R, y_R)$ satisfy $y_R = g(x_R)$, then $DC/V_R = g(CU_R/V_R)$. Hence, the image of the rail on the film (P) is the curve

$$(I): V g(CU/V) = DC. \qquad (2)$$

Now consider a particle $p_E$ with coordinates $(x_E, y_E, C)$ on the edge of the train, which is in contact with the rail. The light ray from $p_E$, at $t = t_E = -\sqrt{x_E^2 + y_E^2 + C^2}/c$, will reach $O$ at $t = 0$ and impress (P) at $(U_E, V_E)$. To resolve the train paradox we have to show that for any particle $p_E$, the point $(U_E, V_E)$ will lie on the curve (I) given by (2). According to the general formula (1), $(U_E, V_E)$ are related to $(x_E, y_E, C)$ by $(x_E, y_E) = (CU_E/V_E, DC/V_E)$. Since $p_E$ is constantly in contact with the rail, at $t_E$ there exists a particle $p_R$ on the rail such that $(x_R, y_R) = (x_E, y_E)$, and hence $(x_R, y_R) = (CU_E/V_E, DC/V_E)$. From the relation $y_R = g(x_R)$, we obtain $DC/V_E = g(CU_E/V_E)$, which means that $(U_E, V_E)$ lies on the curve (I) given by (2).

---

[4] While it is a true effect, the Lorentz contraction is, in this case, completely compensated by the effect of signal delay of light rays emitted from an extended object to reach the location of the film at the same date.

When dealing with the *curvilinear* motion, an alternative *analytical* approach to resolve the apparent paradox by the consideration of successive co-moving inertial frames is possible. By applying the explicit (linear) Lorentz transformations to the motion of the proper referential frame of a particle $p_E$ at the retarded time $t_E$, one can express, in a kind of *non-linear* Lorentz relations, the coordinates $(x_E, y_E)$ in $S$, and hence their projections $(U_E, V_E)$, in terms of the coordinates $(x'_E, y'_E)$ of the same particle $p_E$ in the co-moving inertial frame at the retarded time $t_E$ [7]. The resolution of the paradox results showing that $(U_E, V_E)$ and $(U_R, V_R)$ lie on the same line in (P). This was applied in [1] for the restricted case of a uniformly moving train along a straight rail by a direct application of the relevant analytical transformations as arising from the Lorentz transformations.

To summarize the situation, when photographing an extended object, the longitudinal sections appear to have undergone non-equal and non-proportional displacements [8], parallel to the object's velocity vector. The closer the longitudinal section is to the lens the greater the displacement. As a result, any non-longitudinal section appears to have rotated counterclockwise if the body was moving to the right of the reader.

### 3. The Photographic Process Theorem

In the following, we will give some definitions and establish a theorem on the photographic process of two moving objects. The lens is at rest at the origin, $O$, of the referential of observation, $S$. All the space-time coordinates encountered below are those of $S$. We suppose that the film is exposed to receive light rays at $t = 0$. We consider the case of two extended objects moving with instantaneous velocities $\vec{v}_1(t)$ and $\vec{v}_2(t)$ in the inertial frame $S$. We suppose that the two objects come close enough to each other to be photographed together.

### 3.1. Definitions

- The *Photographic Process* includes all the light emission events that take place during the photographic interval, $I_F$, defined as follows. In order to reach point $O$ at $t = 0$, the light rays emanating from the different points of the two objects were emitted at different negative dates. Let $t_L$ and $t_U$ be the lower and the upper limits of these dates, respectively. We define the *photographic interval* $I_F$ by $I_F = [t_L, t_U] \subset [t_L, 0)$. A further remark in this definition is that the positive value $(-t_L)$ is the time spent by light to travel from point $O$ to the *farthest photographed* point of the two objects when the latter are considered at $t = t_L$; and $(-t_U)$ is the time spent by light to travel from point $O$ to the *nearest photographed* point of the two objects when the latter are considered at $t = t_U$.
- The *Retarded Region of Contact*, $D_R(t)$, is defined as follows. Throughout this letter, the *material points* that built the two objects are called *single points*. At a given time $t \in I_F$, suppose that the single point $^1B(t)$, belonging to object 1, meets the single point $^2B(t)$, belonging to object 2, at some *geometric point* $P$ in space. The doublet of single points $^1B(t)$ and $^2B(t)$ that meet at $P$, at $t$, is called a *couple*, $C(t) = (^1B(t), {}^2B(t))$, and the point $P$ is called a *locus of meeting* or *locus of contact*. The set of all these geometric points $P$ in space, that were the loci of contact of points belonging to the two objects at a *given time* $t \in I_F$, is called the *retarded region of contact*, $D_R(t)$. Notice that if the distance $OP \neq (-ct)$, the two light rays emanating from the couple $C(t)$, starting from the point $P \in D_R(t)$ at $t$, will not reach $O$ at $t = 0$. Of *particular interest* is the sub-set

$^0D_R(t) \subset D_R(t)$ for which light ray emissions at $t$ reach $O$ at $t = 0$; it is necessarily the set of all points $P$ such that $OP = (-ct)$.

- The *Integrated Region of Contact*, $D_I$, is the region in space that is the union of all $D_R(t)$ for $t \in I_F$. A point $P \in D_I$ may be the locus at which several couples meet, such a point is called a *multi-meeting* point. A property of a multi-meeting point is that through it, single points belonging to only one object, may pass too. Consequently, a light ray, starting from a multi-meeting point that could reach $O$ at $t = 0$ may originate from a single point[5]. An explicit example is given in footnote number 7. Conversely, a couple may meet at two different points or more, in $D_I$, at different dates. The important thing in this definition is that, while the points of $D_I$ are not sources for light emissions[6], they are considered in this issue as the *starting points* for light emissions. Since they are *fixed* points in space, the study of light emitted by the couples meeting at these points, or emitted by the single points passing through the same points, and the reception of light by $O$, is greatly simplified.
- The *Permanent Region of Contact*, $D_P$, is the region in space that is the intersection of all $D_R(t)$ for $t \in I_F$, when all these are non-empty sets. A point $P \in D_P$ is called a *permanent locus of meetings*, and it is by definition a multi-meeting point. In the case where some of the sets $D_R(t)$ are empty, the intersection of all the remaining non-empty sets is called a *semi-permanent region of contact*, $D_{SP}$.
- The *Photographed Region*, $D_G$, includes any point in space that is a starting point for light ray emission, emanating from a single point or a couple, that reaches $O$ at $t = 0$.
- The *Retarded Image*, $R_I(t)$, is the set of points on the film that are impressions of light rays emitted by single points or couples at the same retarded date $t \in I_F$. That is, the image of $D_R(t)$ on the film is a sub-set of $R_I(t)$.
- The *History Image*, $H_I$, is the set of all retarded images $R_I(t)$ for $t \in I_F$. It is the printed image of the photographed region, $D_G$.

### 3.2. Theorem

The following assertions are true.

**1.** At a given time $t \in I_F$, if $^0D_R(t) \subset D_R(t)$ is not empty ($^0D_R(t) \neq \Phi$), the points $P \in {^0D_R(t)}$ are spread out on a spherical shell centered at $O$. As time progresses, the spherical shell collapses at the speed of light in vacuum, $c$.

**2.** One, and only one light ray, among those light ray emissions starting from a multi-meeting point $P \in D_I$ for $t \in I_F$, *might* reach $O$ at $t = 0$. Said otherwise, one, and only one couple (or one single point) among those that meet at (or pass through) a multi-meeting point $P \in D_I$ for $t \in I_F$, *might* be photographed.

**3.** If $D_P$ is not empty ($D_P \neq \Phi$), for each $P \in D_P$ there *exists* one, and only one light ray emission, starting from $P$, that *reaches* $O$ at $t = 0$. Said otherwise, there exists one, and only one couple meeting at $P \in D_P$ that is photographed.

**4. a)** $\bigcup_{t_L}^{t_U} {^0D_R(t)} \subset (D_G \cap D_I)$. The union of all $^0D_R(t)$ is denoted by $\bigcup_{t_L}^{t_U} {^0D_R(t)}$.

**b)** If $D_P \neq \Phi$, then $D_P \subset (D_G \cap D_I)$.

---

[5] Here we emphasize the fact that at a given time $t \in I_F$ only couples meet at points belonging to $D_R(t)$, i.e., through each point $P \in D_R(t)$ passes a couple. Later, during the photographic process, other single points or couples may pass through the same point $P \in D_R(t)$. Hence, the light ray, starting from this point $P \in D_R(t)$ that reaches $O$ at $t = 0$, does not necessarily originate from that couple that passed through $P$ at time $t$; this would be the case if $OP = (-ct)$.

[6] The sources are the couples or the single points, whether they produce light or reflect it.

**c)** If $D_P = \Phi$ & $D_{SP} \neq \Phi$, then $D_{SP} \not\subset (D_G \cap D_I)$.

**5.** The photographed region $D_G$ is divided, by assertion 4a), into two disjoint sub-sets, $\bigcup_{t_L}^{t_U} {}^0D_R(t)$ & $\left( D_G \setminus \bigcup_{t_L}^{t_U} {}^0D_R(t) \right)$. The image on the film produced by light rays emanating from each point $P \in \bigcup_{t_L}^{t_U} {}^0D_R(t)$ is a common image of the two single points of the couple that met at $P$, reproducing the contact that took place in nature. The image on the film produced by light rays emanating from each point $P \in \left( D_G \setminus \bigcup_{t_L}^{t_U} {}^0D_R(t) \right)$ is a single image of the single point that passed through $P$, belonging to one or the other object.

**3.3. Proofs & Comments**

The following outlined steps of proofs & comments are numbered in correspondence with the five assertions.

**1.** Since the speed of light in vacuum is invariant under rotations, the light rays emanating from ${}^0D_R(t)$ at $t \in I_F$ will cross the same distance, $-ct$, before reaching $O$ at $t = 0$. This distance decreases as $t$ increases.

**2.** Points $P \in D_I$ being fixed in space, if there is a light ray emission at $t = -OP/c$ starting from a point $P \in D_I$, then any emissions at earlier or later times from the same point will not reach $O$ at $t = 0$. Said otherwise, if it happens that no single point passes through a point $P \in D_I$ at precisely $t = -OP/c$ then no light emission starting from $P$ will reach $O$ at $t = 0$.

**3.** Any point $P \in D_P$ is by definition a point where only couples meet; no single point passes through $P \in D_P$ for $t \in I_F$. The end-points of the photographic interval $I_F$ being fixed, $t_L$ & $t_U$, consider the region $D$ between the two spheres of radii $-ct_L$ and $-ct_U$ centered at $O$, that we may call the equivalent isotropic region for the photographic process. Check that no light ray originated from a point $P \notin D$, for $t \in I_F$, reaches $O$ at $t = 0$.

a) Suppose that $D_P \subset D$, since there is always a couple meeting at $P \in D_P$ any date $t \in I_F$, check that the light ray emitted from $P \in D_P$ at $t = -OP/c \in I_F$ reaches $O$ at $t = 0$.

b) Suppose that $D_P \not\subset D$, there exists at least $P \in D_P$ and $P \notin D$ for which $OP > -ct_L$ or $OP < -ct_U$. Consider the case $OP > -ct_L$ (the other case is treated similarly), since $P \in D_P$ is a permanent locus of meeting, a couple $C_L$ certainly met there at $t = t_L$. By definition of $t_L$, the time spent by light to travel from point $O$ to any single point or couple of the two objects, when the latter are considered at $t = t_L$, is less than or equal to $(-t_L)$. Since this is not the case with $P \in D_P$, by which the couple $C_L$ passed at $t = t_L$, for $OP/c > (-t_L)$, this rejects the hypothesis $D_P \not\subset D$.

**4. a)** The assertion is obvious since each ${}^0D_R(t)$ is in $D_G$ and $D_I$ at the same time. By definition, any light ray emitted from $P \in \bigcup_{t_L}^{t_U} {}^0D_R(t)$ that reaches $O$ at $t = 0$ originated from a couple that meets there; light rays emitted from single points that might pass through $P \in \bigcup_{t_L}^{t_U} {}^0D_R(t)$ do not reach $O$ at $t = 0$. Said otherwise, if $P \in D_I$ is such that at $t = -OP/c \in I_F$ a light ray is emitted from a single point[7] passing through $P$, then by definition $P \notin \bigcup_{t_L}^{t_U} {}^0 D_R(t)$. Since the light ray is emitted at $t = -OP/c$ it will reach $O$ at $t = 0$, this accounts for $P \in D_G$. Consequently $P \in (D_G \cap D_I)$. Hence, we conclude that $(D_G \cap D_I) \not\subset \bigcup_{t_L}^{t_U} {}^0 D_R(t)$.

**b)** Similar arguments to those given in a) lead to $D_P \subset (D_G \cap D_I)$ & $(D_G \cap D_I) \not\subset D_P$.

**c)** Since a point in $D_R(t)$, and hence in $D_{SP}$, is not necessarily a starting point for light ray emission that reaches $O$ at $t = 0$. In fact, a point $P \in D_{SP}$ is the locus where couples meet or through it, single points pass at some precise dates in $I_F$. If all these dates are, each, different from $-OP/c \in I_F$, such a point is not in $D_G$, and hence it is not in $(D_G \cap D_I)$ too. Compare with [10].

**5.** One may think that the best way to divide the photographed region $D_G$ into two disjoint sub-sets is to consider $(D_G \cap D_I)$ & $\left( D_G \setminus (D_G \cap D_I) \right)$. But this is not the case since $(D_G \cap D_I) \neq \bigcup_{t_L}^{t_U} {}^0 D_R(t)$. The speed of light being independent of the speed of the source, the two light rays emitted from the two single points of a couple meeting at $P \in \bigcup_{t_L}^{t_U} {}^0 D_R(t)$ will travel together in the same direction and impress the film at the same point-image $I \in H_I$ at the same date, here, $t = 0$.

## 4. Conclusion

Our main aim has been achieved by first resolving the train paradox in the general case of curvilinear motion with varying speed. Since the idea behind the resolution is based on the application of two absolute facts or postulates of Special Relativity itself, hence the paradox in question is not a true one. A true paradox is one that emerges from an inconsistency of the theory applied. We have emphasized that the future motion of the light ray that started from a point $P$ in space at time $t$, its speed and destination are independent of every thing, intrinsic, extrinsic or whatever property, related with the source of the light ray which passed through $P$ at $t$, whenever the propagation of the light ray is in vacuum. The way we have proceeded required us to formulate, in a second step, a general theorem on the topology of sets of fixed points that are the starting points for light ray emissions during a given time interval ($I_F$) to reach a given point ($O$) at a 'given date' ($t = 0$), and on the time relationships between these sets and the given point. Because of the homogeneity of the time axis, the 'given date' may be chosen arbitrarily; hence the theorem depends only on five interdependent parameters, namely the space coordinates of the given point and the endpoint of the time interval. The order in which our account proceeded is constructive and illustrative. Reviewing the account in reverse order, the resolution of the paradox appears as a special case, say an application, of the theorem.

The theorem remains valid in case of many objects but needs some generalization. For instance, if three objects were to be photographed, we will still have points where only couples meet, points where only triplets meet, and points where couples and triplets meet, as well as, single points. Hence one needs to extend the definition of a meeting point to include triplets, as well as the definitions of some sets, in addition to establishing other new definitions. This is an open problem.

---

[7] To see how a light ray that reaches $O$ at $t = 0$ may originate from a single point that passes through $P \in D_I$, consider a point $P$ on a sheet of paper, a long with $O$ such that $OP$ is too large, and bring the head of a needle too close to $P$. The light ray emitted from $P$ (single point) at $t = -OP/c$ reaches $O$ at $t = 0$, so $P \in D_G$. Before this light ray ever reaches $O$, let the head of the needle touch $P$ (couple) at some $t \in I_F$ & $t > -OP/c$, hence $P \in D_I$. It is clear that $P$ is not in one of the sets ${}^0 D_R(t)$ for the light ray emitted from $P$ at the time when the needle touches the sheet, or later, will not reach $O$ at $t = 0$.

A useful set of problems and exercises for an undergraduate course level can be derived from the theorem [9], as a guideline to students to better understand the applications of Special Relativity without calculus and algebra [10]. Applications of the theorem to Astronomy, if any, would be a further step.

**Acknowledgments**

The author expresses his gratitude to Eddie J. Girdner for revising the manuscript, and thanks one the Referees for his critical remarks.